\documentclass[twocolumn,showpacs,amsmath,amssymb,aps,prb]{revtex4}
\usepackage{graphicx}

\begin{document}
\def\kag{{\it kagom\'e }}
\def\et{{\it et al.}}

%% Title and Abstract
\title{Numerical Contractor Renormalization Method for Quantum Spin Models}
\author{Sylvain Capponi}
\email{capponi@irsamc.ups-tlse.fr}
\author{Andreas L\"auchli}
\author{Matthieu Mambrini}
\affiliation{
  Laboratoire de Physique Th\'eorique, CNRS UMR 5152,
  Universit\'e Paul Sabatier, F-31062 Toulouse, France
}

\begin{abstract}
  We demonstrate the utility of the numerical Contractor Renormalization (CORE) 
  method for quantum spin systems by studying one and two dimensional model cases.
  Our approach consists of two steps:
  (i) building an effective Hamiltonian with longer ranged interactions up to a certain cut-off
  using the CORE algorithm and
  (ii) solving this new model numerically on finite clusters by exact diagonalization and 
  performing finite-size extrapolations to obtain results in the thermodynamic limit. 
  This approach, giving complementary information to analytical treatments of the CORE 
  Hamiltonian, can be used as a semi-quantitative numerical method.
  For ladder type geometries, we explicitely check the accuracy of the effective models by
  increasing the range of the effective interactions until reaching convergence.
  Our results in the perturbative regime and also away from it are in good agreement with 
  previously established results.
  In two dimensions we consider the plaquette lattice and the \kag lattice as non-trivial
  test cases for the numerical CORE method. As it becomes more difficult to extend the range 
  of the effective interactions in two dimensions, we propose diagnostic tools (such
  as the density matrix of the local building block) to ascertain the validity of the basis
  truncation. On the plaquette lattice we have an excellent description 
  of the system in both the disordered and the ordered phases, thereby showing that the CORE
  method is able to resolve quantum phase transitions. On the \kag lattice we find that the
  previously proposed twofold degenerate $S=1/2$ basis can account for a large number of 
  phenomena of the spin $1/2$ \kag system. For spin $3/2$ however this basis does not seem 
  to be sufficient anymore.
  In general we are able to simulate system sizes which correspond to an $8\times 8$ lattice for the
  plaquette lattice or a 48-site \kag lattice, which are beyond the possibilities of a
  standard exact diagonalization approach.
\end{abstract}
\date{\today}
\maketitle

% I. Introduction
Low-dimensional quantum magnets are at the heart of current interest in strongly 
correlated electron systems. These systems are driven by strong correlations 
and large quantum fluctuations - especially when frustration comes into play -
and can exhibit various unconventional phases and quantum phase transitions.

One of the major difficulties in trying to understand these systems is
that strong correlations often generate highly non trivial low-energy
physics. Not only the groundstate of such models is generally not known
but also the low-energy degrees of freedom can not be identified easily.
Moreover, among the techniques available to investigate these systems,
not many have the required level of generality to provide a systematic
way to derive low-energy effective Hamiltonians.

%%In order to fully understand these systems powerful analytical and numerical methods are
%%needed. 
Recently the Contractor Renormalization (CORE) method has been introduced
by Morningstar and Weinstein \cite{weinstein96}. The key idea of the approach is
to derive an effective Hamiltonian acting on a truncated local basis set, so as
to exactly reproduce the low energy spectrum. In principle the method is exact 
in the low energy subspace, but only at the expense of having a priori long
range interactions. The method becomes most useful when one can significantly 
truncate a local basis set and still restrict oneself to short range effective
interactions. This however depends on the system under consideration and has to
be checked systematically.
Since its inception the CORE method has been mostly used as an analytical method 
to study strongly correlated systems \cite{weinstein01,core-bosonfermion,core-pyrochlore}. 
Some first steps in using the CORE approach and related ideas in a numerical framework have
also been undertaken \cite{shepard97,dynamicCORE,core-tJladder,malrieu}.

The purpose of the present paper is to explore the numerical CORE method as a 
complementary approach to more analytical CORE procedures, and to systematically 
discuss its performance in a variety of low dimensional quantum magnets, both 
frustrated and unfrustrated. The approach consists basically of numerical exact
diagonalizations of the effective Hamiltonians. In this way a large number of 
interesting quantities are accessible, which otherwise would be hard to 
obtain. Furthermore we discuss some criteria and tools useful to estimate the
quality of the CORE approach. 

The outline of the paper is as follows:
In the first section we will review the CORE algorithm in general and discuss some 
particularities in a numerical CORE approach, both at the level of the calculation
of the effective Hamiltonians and the subsequent simulations.

In section~\ref{sec:ladder} we move to the first applications on one-dimensional (1D) systems:
the well known two-leg spin ladder and the 3-leg spin ladder with periodic boundary conditions
in the transverse direction (3-leg torus). Both systems exhibit generically
a finite spin gap and a finite magnetic correlation length. We will show that the numerical
CORE method is able to get rather accurate estimates of the groundstate energy and the
spin gap by successively increasing the range of the effective interactions.

In section~\ref{sec:2D} we discuss two-dimensional (2D) systems. As in 2D a long ranged
cluster expansion of the interactions is difficult to achieve on small clusters, we will 
discuss some techniques to analyze the quality of the basis truncation. We illustrate these issues
on two model systems, the plaquette lattice and the \kag lattice.
The plaquette lattice is of particular interest as it exhibits a quantum phase transition from a 
disordered plaquette state to a long range ordered N\'eel antiferromagnet, which cannot be reached by 
a perturbative approach. We show that a range-two effective model captures many aspects
of the physics over the whole range of parameters.
The \kag lattice on the other hand is a highly frustrated lattice built of corner-sharing triangles.
For spin $1/2$ it has been studied both numerically and analytically and it is one best-known candidate
systems for a spin liquid groundstate. A very peculiar property is the exponentially large number of
low-energy singlets in the magnetic gap. We show that already a basic range two CORE approach is able
to devise an effective model which exhibits the same exotic low-energy physics. For higher half-integer
spin, i.e. $S=3/2$, this simple effective Hamiltonian breaks down; we analyze how to detect this, and
discuss some ways to improve the results.

In the last section we conclude and give some perspectives. Finally three appendices 
are devoted to (i) the density matrix of local building block, (ii) the calculation of observables by
energy considerations and (iii) some general remarks on effective Hamiltonians coupling 
antiferromagnetic half-integer spin triangles.

\section{CORE Algorithm}

The Contractor Renormalization (CORE) method has been proposed by Morningstar and Weinstein in the
context of general Hamiltonian lattice models~\cite{weinstein96}.
Later Weinstein applied this method with success to various spin chain models~\cite{weinstein01}.
For a review of the method we refer the reader to these original papers~\cite{weinstein96,weinstein01}
and also to a pedagogical article by Altman and Auerbach~\cite{core-bosonfermion} which includes
many details. Here, we summarize the basic steps before discussing some technical aspects
which are relevant in our numerical approach.

{\it CORE Algorithm~:}
\begin{itemize}
\item 
  Choose a small cluster (e.g. rung, plaquette, triangle, etc) and
  diagonalize it. Keep $M$ suitably chosen low-energy states. 
\item
  Diagonalize the full Hamiltonian $H$ on a connected graph 
  consisting of $N_c$ clusters and obtain its low-energy states $|n\rangle$
  with energies $\varepsilon_n$.
\item
  The eigenstates $|n\rangle$ are projected on the tensor product
  space of the states kept
  and Gram-Schmidt orthonormalized in order to get a basis
  $|\psi_n\rangle$ of dimension $M^{N_c}$.
  As it may happen that some of the eigenstates have zero or
  very small projection, or vanish after the orthogonalization
  it might be necessary to obtain more than just $M^{N_c}$ exact 
  eigenstates.
\item
  Next, the effective Hamiltonian for this graph is built as
  \begin{equation}
    h_{N_c} = \sum_{n=1}^{M^{N_c}} \varepsilon_n
    |\psi_n\rangle\langle \psi_n|.
  \end{equation}
\item 
  The connected range-$N_c$ interactions $h^{\rm conn}_{N_c}$ are determined by
  substracting the contributions of all connected subclusters. 
\item
  Finally, the effective Hamiltonian is given by a cluster expansion
  as
  \begin{equation}
    H^{\mbox{\tiny CORE}}=\sum_i h_i +\sum_{\langle
      ij\rangle}h_{ij}+\sum_{\langle ijk\rangle} h_{ijk} +\cdots
  \end{equation}
\end{itemize}

This effective Hamiltonian \emph{exactly} reproduces the low-energy 
physics provided the expansion goes to infinity. However, if the interactions 
are short-range in the starting Hamiltonian,
we can expect that these operators will become smaller and smaller, at least
in certain situations. 
In the following, we will truncate at range $r$ and verify the convergence in 
several cases. This convergence naturally depends on the number $M$ of low-lying
states that are kept on a basic block. In order to describe quantitatively how 
``good'' these states are, we introduce the density matrix in section~\ref{sec:2D}.

When the number of blocks increases, a full diagonalization is not
always easy and one is tempted to use a Lanczos algorithm in order to
compute the low-lying eigenstates. In that case, one has to be very
careful to resolve the correct degeneracies, which is known to be a difficult
task in the Lanczos framework. 
In practice such degeneracies arise when the cluster to be diagonalized
is highly symmetric. If the degeneracies are ignored, often a wrong effective
Hamiltonian with broken SU(2) symmetry is obtained. As a consequence we 
recommend to use specialized \verb:LAPACK: routines whenever possible.

In the present work we investigate mainly SU(2) invariant Heisenberg models described by the usual 
Hamiltonian
\begin{equation}
H=\sum_{\langle ij\rangle} J_{ij} \vec{S}_i \cdot \vec{S}_j
\end{equation}
where the exchange constants $J_{ij}$ will be limited to short-range distances
in the following.  
As a consequence of the SU(2) symmetry, the total spin of all states is a good 
quantum number. This also
has some effects when calculating the effective Hamiltonian. It is possible 
to have situations where a low energy state has a non-zero overlap
with the tensor product basis, but gets eliminated by the orthogonalization 
procedure because one has already exhausted all the states in one particular
total spin sector by projecting states with lower energy.

Once an effective Hamiltonian has been obtained, it is still
a formidable task to determine its properties. Within the CORE
method different routes have been taken in the past. In their 
pioneering papers Morningstar and Weinstein have chosen to iteratively 
apply the CORE method on the preceding effective Hamiltonian
in order to flow to a fixed point and then to analyze the fixed
point. A different approach has been taken in Refs. 
[\onlinecite{core-bosonfermion,core-pyrochlore}]: There the 
effective Hamiltonian after one or two iterations has been
analyzed with mean-field like methods and interesting results
have been obtained. Yet another approach - and the one we will
pursue in this paper - consists of a single CORE step to
obtain the effective Hamiltonian, followed by a numerical simulation
thereof. This approach has been explored in a few previous studies
\cite{shepard97,dynamicCORE,core-tJladder}. The numerical technique we employ
is the Exact Diagonalization (ED) method based on the Lanczos 
algorithm. This technique has easily access to many observables
and profits from the symmetries and conservation laws in the 
problem, i.e. total momentum and the total $S^z$ component. Using 
a parallelized program we can treat matrix problems of dimensions up
to $\sim 50$~millions, however the matrices contain significantly more matrix 
elements than the ones of the microscopic Hamiltonian we start with.

% II. Ladder Geometries
\section{Ladder geometries}\label{sec:ladder}
In this section, we describe results obtained on ladder systems with 
2 and 3 legs respectively. 

\begin{figure}
  \centerline{\includegraphics*[width=0.9\linewidth]{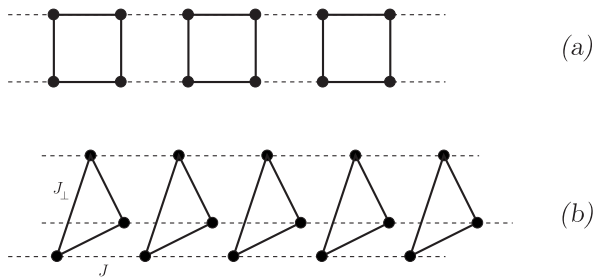}}
  \caption{
    (a) 2-leg ladder. Basic block is a $2\times 2$ plaquette.
    (b) 3-leg torus with rung coupling $J_\perp$ and inter-rung coupling 
$J_\parallel$.
    \label{fig:Lattices1D}
  }
\end{figure}

We want to build an effective model that is valid from a 
perturbative regime to the isotropic case $J_{ij}=J=1$.
We have chosen periodic boundary conditions
(PBC) along the chains in order to improve the convergence to the thermodynamic limit. 

\subsection{Two-leg Heisenberg ladder}
\label{sec:LadderGeometries}
The 2-leg Heisenberg ladder has been intensively studied and is known 
to exhibit a spin gap for all couplings~\cite{2leg,2legdata}. 

In order to apply our algorithm, we select a $2\times 2$ plaquette
as the basic unit (see Fig.~\ref{fig:Lattices1D} (a)). The truncated subspace is 
formed by the singlet ground-state (GS) and the lowest triplet state.

Using the same CORE approach, 
Piekarewicz and Shepard have shown that quantitative results can be obtained 
within this restricted subspace~\cite{shepard97}. Moreover, dynamical quantities 
can also be computed in this framework~\cite{dynamicCORE}.

\begin{figure}
  \includegraphics*[width=\linewidth]{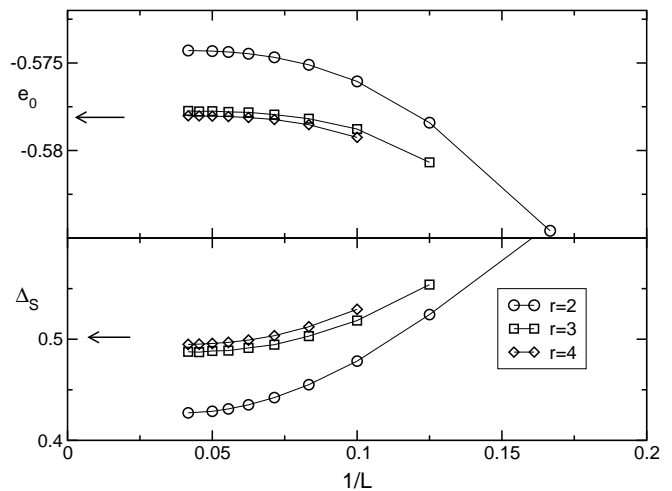}
  \caption{
    Ground-state energy per site and spin
    gap  of a $2\times L$ Heisenberg ladder using CORE method with
    various range $r$ using PBC. 
    For comparison, we plot the best known extrapolations~\protect{\cite{2legdata}} with arrows.
    \label{fig:heisen_2xL}
  }
\end{figure}

Since we are dealing with a simple system, we can compute the effective 
models including rather long-range interactions (typically, to obtain 
range-4 interactions, we need to compute the low-lying states on a 
$2\times 8$ lattice with Open Boundary Conditions which is feasible, although it requires 
a large numerical effort). It is desirable to compute long-range 
effective interactions since we wish to check how the truncation affect 
the physical results and how the convergence is reached. 

In a second step, for each of these effective models, we perform a 
standard Exact Diagonalization (ED) using the Lanczos algorithm on finite clusters up to $N_c=12$ 
clusters ($N=48$ sites for the original model).
The GS energy and the spin gap are shown in Fig.~\ref{fig:heisen_2xL}. 
The use of PBC allows to reduce considerably
finite-size effects since we have an exponential convergence as a
function of inverse length.  
CORE results are in perfect agreement with known results and 
the successive approximations converge uniformly to
the exact results. For instance, the relative errors of range-4 results are 
$10^{-4}$ for the GS energy and $10^{-2}$ for the spin gap. This fast convergence is probably 
due to the rather short correlation length in an isotropic ladder (typically 3 to 4 lattice 
spacings~\cite{xi_2leg}). 

\subsection{3-leg Heisenberg torus}
As a second example of ladder geometry, we have studied 
a 3-leg Heisenberg ladder with PBC along the rungs. This property 
causes geometric frustration which leads to a finite
spin-gap and finite dimerization for all interchain coupling $J_\perp$~\cite{kawano97,cabra98},
contrary to the open boundary condition case along the rungs, which is in the 
universality class of the Heisenberg chain.

{\it Perturbation theory~:} 
The simple perturbation theory is valid when the coupling along the
rung ($J_\perp$) is much larger than between adjacent rungs ($J_\parallel$). 
In the following, we fix $J_\perp=1$ as the energy unit and denote $\alpha=J_\parallel/J_\perp$. 

On a single rung, the low-energy states are the following
degenerate states, defined as
\begin{eqnarray}
 \mid\uparrow L\rangle&=&\frac{1}{\sqrt{3}}(\mid \uparrow \uparrow \downarrow
\rangle +\omega \mid\uparrow \downarrow \uparrow \rangle +\omega^2
\mid\downarrow \uparrow \uparrow \rangle), \label{eqn:TorusChirality} \\
\mid\downarrow L\rangle&=&\frac{1}{\sqrt{3}}(\mid\downarrow \downarrow
\uparrow \rangle +\omega \mid\downarrow \uparrow \downarrow \rangle
+\omega^2 \mid\uparrow \downarrow \downarrow \rangle), \nonumber \\
\mid\uparrow R\rangle&=&\frac{1}{\sqrt{3}}(\mid\uparrow \uparrow
\downarrow \rangle +\omega^2 \mid\uparrow \downarrow \uparrow \rangle
+\omega
\mid\downarrow \uparrow \uparrow \rangle),  \nonumber\\
\mid\downarrow R\rangle&=&\frac{1}{\sqrt{3}}(\mid\downarrow \downarrow
\uparrow \rangle +\omega^2 \mid\downarrow \uparrow \downarrow \rangle
+\omega \mid\uparrow \downarrow \downarrow \rangle) 
\nonumber
\end{eqnarray}
 where
$\omega=\exp(i2\pi/3)$. The indices $L$ and $R$ represent the
momentum of the 3-site ring $k_y=2\pi/3$ and $-2\pi/3$ respectively.
They define two chiral states which can be viewed as pseudo-spin
states with operators $\vec{\tau}$ on each rung defined by
\begin{eqnarray*}
\tau^+ \mid\cdot\, R\rangle = 0 \qquad \tau^+ \mid \cdot\, L\rangle = | \cdot\, R\rangle\\ 
\tau^- \mid \cdot\, R\rangle = | \cdot\, L\rangle \qquad \tau^- \mid \cdot\, L\rangle = 0 \\
\tau^z \mid \cdot\, R\rangle = \frac{1}{2} \mid \cdot\, R\rangle \quad \tau^z \mid \cdot\, L\rangle = -\frac{1}{2} \mid \cdot\, L\rangle 
\end{eqnarray*}
 These states have in addition a physical spin 
1/2 described by $\vec{\sigma}$.

Applying the usual perturbation theory for the inter-rung coupling, one
finds\cite{schulz,kawano97} 
\begin{equation}\label{H3leg_pert}
H_{pert}=-\frac{N}{4}+\frac{\alpha}{3}\sum_{\langle i j \rangle}
\vec{\sigma}_i\cdot \vec{\sigma}_j (1+4 (\tau_i^+ \tau_j^- + \tau_i^-
\tau_j^+))
\end{equation}
where $N$ is the total number of sites. 

This effective Hamiltonian has been studied with DMRG and ED
techniques and it exhibits a finite spin
gap $\Delta_S=0.28~J_\parallel$ and a dimerization of the ground 
state~\cite{kawano97,cabra98}.

Here we want to use the CORE method to extend the perturbative
Hamiltonian with an effective Hamiltonian in the same basis
for \emph{any} coupling.

{\it CORE approach~:} 
As a basic unit, we choose a single 3-site rung. 
The subspace consists of the same low-energy states as for the perturbative result 
(Eq.~(\ref{eqn:TorusChirality})) 
which are 4-fold degenerate (2 degenerate $S=1/2$ states). We can
apply our procedure to compute the effective interactions at various ranges, 
in order to be able to test  the convergence of the method.

First, we write down the range-2 contribution under the most
general form which preserves both $SU(2)$ (spin) symmetry and 
simultaneous translation 
or reflection along all the rungs~:
\begin{eqnarray}
H_{r=2}=N a_0 + \sum_{\langle i j \rangle}(b_0 \tau_i^z \tau_j^z
+c_0 (\tau_i^+ \tau_j^- + \tau_i^- \tau_j^+)) \nonumber\\
 + \vec{\sigma}_i\cdot
\vec{\sigma}_j (a_1+  b_1 \tau_i^z \tau_j^z+ c_1(\tau_i^+ \tau_j^- +
\tau_i^- \tau_j^+))\label{heff_3xL}
\end{eqnarray}

In the perturbative regime given in (\ref{H3leg_pert}), the only non-vanishing 
coefficients are given by~: $a_0=-1/4$, $a_1=\alpha/3$ and $c_1=4\alpha/3$. 

\begin{figure}[h]
\includegraphics*[width=0.9\linewidth]{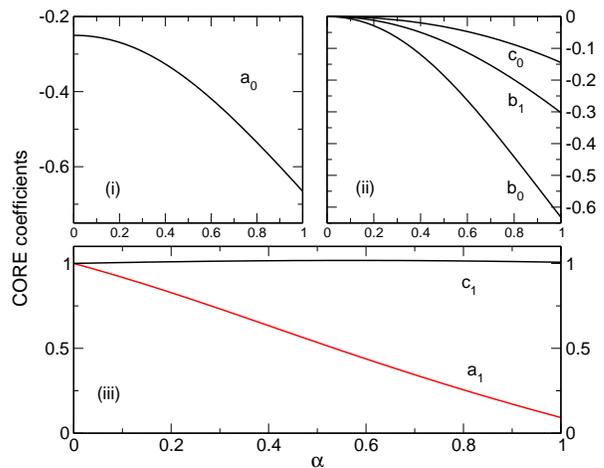}
\caption{CORE coefficients (see Eq.~\ref{heff_3xL}) for two coupled triangles as a function of 
the inter-rung coupling $\alpha=J_\parallel/J_\perp$. 
The parameters were computed using range-2 CORE.  The coefficients in panel (iii) have been divided by their
    values in the perturbative limit. They therefore all start at 1.} \label{fig:3leg_parameter}
\end{figure}

The parameters of the effective Hamiltonian can be obtained and their dependence as a function of 
the inter-rung coupling $\alpha$ is shown in Fig.~\ref{fig:3leg_parameter}. 
We immediately see some deviations from the perturbative result since coefficients in panel (i) and 
(ii) are non-zero 
and become as important as the other terms in the isotropic limit. Surprisingly, we observe that $c_1$ 
follows its perturbative expression on the whole range of couplings whereas $a_1$ deviates strongly as 
one goes to the isotropic case but does not change sign.

In order to study how the physical properties evolve as a function of $J_\parallel/J_\perp$, we
 have computed the GS energy 
and the spin gap both for a small-coupling case and in the isotropic limit, up to range 5 in the 
effective interactions. 

\begin{figure}[h]
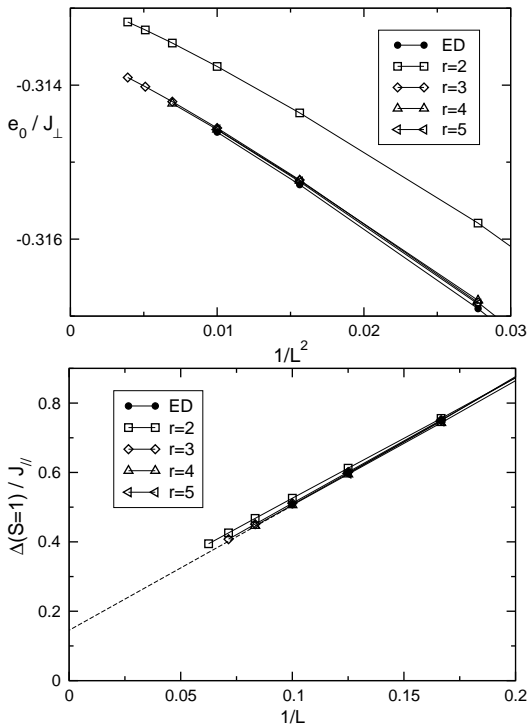

\includegraphics*[width=0.8\linewidth]{heisen_3xL_e0_Jp4}

\includegraphics*[width=0.8\linewidth]{heisen_3xL_gap_Jp4}
\caption{GS energy per site and spin gap for a  $3\times L$
  Heisenberg torus with $J_{\parallel}/J_\perp=0.25$. Results are obtained using the
  CORE method at various range $r$. 
  \label{fig:heisen_3xL_Jp4}
}
\end{figure}

{\it Small interrung coupling~:}
We have chosen $J_\parallel/J_\perp=0.25$ which corresponds to a case 
where perturbation theory should still apply. Using ED, we can solve the 
effective models on finite lattices and on Fig.~\ref{fig:heisen_3xL_Jp4}, we plot 
the scaling of the GS energy and of the spin gap as a function of the system length $L$.
Even for this rather small value of $J_\parallel/J_\perp$, our effective Hamiltonian can be considered
as an improvement over the first order perturbation theory. Moreover, we observe a fast 
convergence with the range of interactions and already the range-3 approximation 
is almost indistinguishable from ED results. 

The estimated gap is $0.16 J_\parallel$ and correspond to a lower bound since ultimately the 
gap should converge exponentially to its thermodynamic value. Our value is consistent with the DMRG 
one~\cite{kawano97} ($\sim 0.2  J_\parallel$), and is 
already reduced compared to the strong coupling result~\cite{kawano97} ($\Delta_S=0.28 J_\parallel$).

\begin{figure}[h]
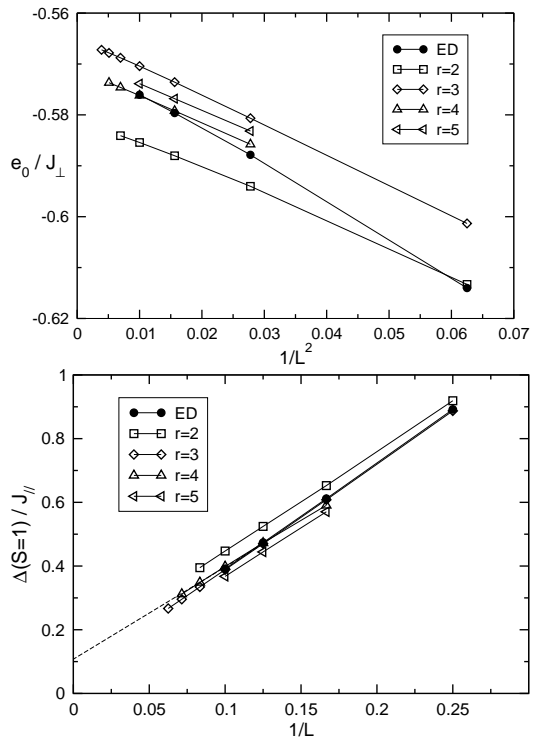

\includegraphics*[width=0.8\linewidth]{heisen_3xL_e0}

\includegraphics*[width=0.8\linewidth]{heisen_3xL_gap}
\caption{Same as Fig.~\ref{fig:heisen_3xL_Jp4} for the isotropic case 
$J_{\parallel}=J_{\perp}=1$.} \label{fig:heisen_3xL}
\end{figure}

{\it Isotropic case~:}
We apply the same procedure in the isotropic limit. 
As expected, the convergence with 
the range of interactions is much slower  than in the perturbative regime.
We show on Fig.~\ref{fig:heisen_3xL} that indeed the
ground state energy converges slowly and oscillates around the correct
value. These oscillations come from the fact that, in order to compute range-$r$ 
interactions, one has to study alternatively clusters with an even or odd number of sites. 
Since this system has a tendency to form dimers on nearest-neighbour bonds, it is better to 
compute clusters with an even number of sites. 

For the spin gap, we find accurate results even with
limited range interactions. In particular, we find that frustration induces a finite 
spin gap $\simeq  0.11~J_\parallel$ in that system. As in the previous case, this is a lower bound which is in
perfect agreement with DMRG study~\cite{kawano97}.

Moreover, we find that the singlet gap vanishes in the thermodynamic limit 
as $1/L^2$ (data not shown). This singlet state at momentum $\pi$ along
the chains corresponds to the state 
built in the generalized Lieb-Schultz-Mattis argument~\cite{LSM}. Here, the physical
picture is a two-fold degenerate GS due to the appearance of spontaneous dimerization.

{\it Spinon dispersion relation for the spin tube~:}
One of the advantages of this method is to be able to get
information on some quantum numbers (number of particles,
magnetization, momentum\ldots) For example, the effective Hamiltonian $H_{\rm eff}$ still
commutes with translations along the legs, with the total
$S^{tot}_z$ and $\tau_z$ so that we can work in a given momentum
sector $(k_x,k_y)$ with a fixed magnetization $S^{tot}_z$. By
computing the energy in each sector,  we can compute the
dispersion relation.

\begin{figure}[h]

\includegraphics*[width=0.8\linewidth]{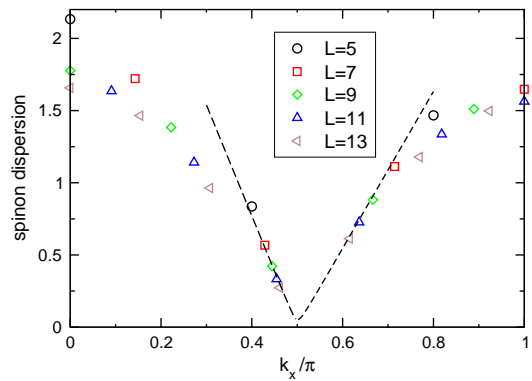}
\caption{Spinon dispersion relation (see text) as a function of
longitudinal momentum (in units of $\pi$). We only plot the 
lowest branch corresponding to  $k_y=\pm 2\pi/3$. The odd lengths
run from 5 to 13. The lines are guide to the eyes for an extrapolation 
on both sides of
$\pi/2$.} \label{fig:spinon}
\end{figure}

In order to try to identify if the fundamental excitation is a spinon, we 
compute the energy difference between the lowest $S=1/2$ state when the length is odd 
($L=2p+1$) and the extrapolated GS energy obtained from the data on systems 
with even length $2p$ and $2p+2$. The data are taken from CORE
with range-4 approximation. On  Fig.~\ref{fig:spinon}, we plot this 
dispersion as a function of the longitudinal momentum, relative 
to the GS with $L=2p$. 

We observe a dispersion compatible with a spinon-like dispersion, 
  which is massive with a gap at
$\pi/2$ $\simeq 0.05 \simeq \Delta_S/2$. This result is consistent 
with a picture in which the triplet excitation $\Delta_S$ is made of 
two elementary spinons. With our precision, it seems that the
spinons are not bound but  we cannot
exclude a small binding energy.

We have on overall agreement with results obtained in the strong
interchain coupling regime~\cite{cabra98}.

Therefore, with CORE method, we have both the advantage of working
in the reduced subspace and not being limited to the perturbative
regime. Amazingly, we have observed that for a very small effort (solving a small cluster), the effective 
Hamiltonian gives much better results (often less than 1\% on GS energies) than perturbation theory. 
It also gives an easier framework to systematically improve the accuracy by including longer range
interactions. 

For these models, the good convergence 
of CORE results may be due to the fact that the GS in the isotropic limit is 
adiabatically connected to the perturbative one. In the following part we will 
therefore study 2D models where a quantum phase transition occurs as one goes from
the perturbative to the isotropic regime.

\section{Two dimensional spin models}
\label{sec:2D}

In this section we would like to discuss the application of the
numerical CORE method to two dimensional quantum spin systems. We 
will present spectra and observables and also discuss a novel 
diagnostic tool - the density matrix of local objects - in order to
justify the truncation of the local state set.

One major problem in two dimension is the more elaborate cluster
expansion appearing in the CORE procedure. Especially our
approach based on numerical diagonalization of the resulting CORE
Hamiltonian faces problems once the CORE interaction clusters wrap
around the boundary of the finite size clusters. We therefore try
to keep the range of the interactions minimal, but we still demand a
reasonable description of low energy properties of the system.
We will therefore discuss some ways to detect under what circumstances
the low-range approximations fail and why.

\begin{figure}
  \centerline{\includegraphics*[width=0.9\linewidth]{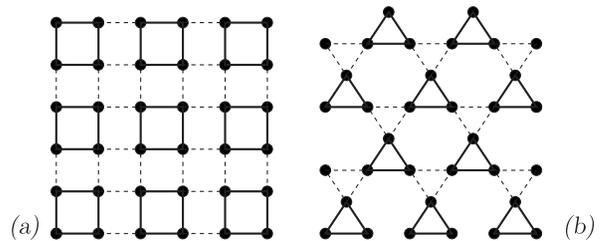}}
  \caption{
    (a) The plaquette lattice. Full lines denote the plaquette bonds $J$, 
    dashed lines denote the inter-plaquette coupling $J'$.
    (b) The trimerized \kag lattice. Full lines denote the up-triangle
    $J$ bonds, dashed lines denote the down-triangle coupling $J'$.
    The standard \kag lattice is recovered for $J'/J=1$.
    \label{fig:Lattices2D}
  }
\end{figure}

As a first example we discuss the plaquette lattice [Fig.~\ref{fig:Lattices2D} (a)],
which exhibits a quantum phase transition from a gapped plaquette-singlet state with
only short ranged order to a  long range ordered antiferromagnetic state as a 
function of the interplaquette coupling
\cite{KogaSigmaModelPlaquette,KogaSeriesPlaquette,AMLShastry,VoigtED}.
We will show that the CORE method works particularly well for this model
by presenting results for the excitation spectra and the order parameter.
It is also a nice example of an application where the CORE method is 
able to correctly describe a quantum phase transition, thus going beyond
an augmented perturbation scheme.

The second test case is the highly frustrated \kag lattice 
[Fig.~\ref{fig:Lattices2D} (b)] with non-integer spin, which has been 
intensively studied for $S=1/2$ during the last few years 
\cite{LeungElser,Lecheminant,Waldtmann,MilaPRL,MambriniMila}.
Its properties are still not entirely understood, but some of the features
are well accepted by now: There is no simple local order parameter detectable, 
neither spin order nor valence bond crystal order. There is probably a small spin
gap present and most strikingly an exponentially growing number of low energy singlets 
emerges below the spin gap.
We will discuss a convenient CORE basis truncation which has emerged from a 
perturbative point of view~\cite{Subrahmanyam,MilaPRL,Raghu} and consider an extension
of this basis for higher non-integer spin.

\subsection{Plaquette lattice}

The CORE approach starts by choosing a suitable decomposition
of the lattice and a subsequent local basis truncation. In the
plaquette lattice the natural decomposition is directly given by 
the uncoupled plaquettes. Among the 16 states of an isolated plaquette
we retain the lowest singlet [$K=(0,0)$] and the lowest triplet
[$K=(\pi,\pi)$]. The standard argument for keeping these states
relies on the fact that they are the lowest energy states in the 
spectrum of an isolated plaquette. 

\begin{figure}
  \centerline{\includegraphics*[width=0.9\linewidth]{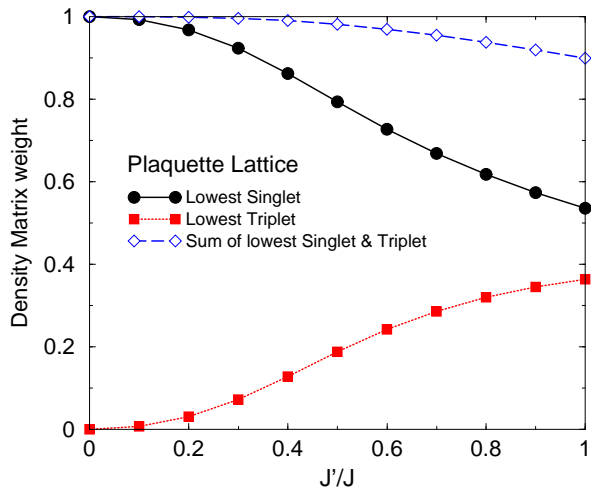}}
  \caption{
    Density matrix weights of the two most important states on
    a strong ($J$-bonds) plaquette as a function of $J'/J$.
    These results were obtained by ED with the original Hamiltonian
    on a $4\times4$ cluster.
    \label{fig:PlaquetteDensityMatrix}
  }
\end{figure}

As discussed in appendix \ref{sec:DensityMatrix}, the density matrix 
of a plaquette in the fully interacting system gives clear indications
whether the basis is suitably chosen. In Fig.~\ref{fig:PlaquetteDensityMatrix}
we show the evolution of the density matrix weights of the lowest
singlet and triplet as a function of the interplaquette coupling. Even
though the individual weights change significantly, the sum of both
contributions remains above 90\% for all $J'/J\le 1$. We therefore
consider this a suitable choice for a successful CORE application.

\begin{figure}
  \centerline{\includegraphics*[width=0.9\linewidth]{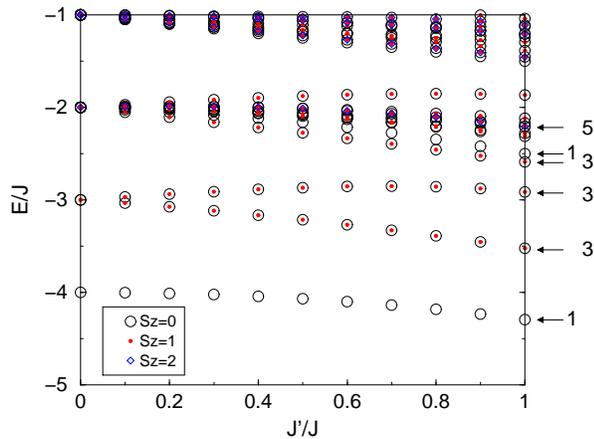}}
  \caption{
    Low energy spectrum of two coupled plaquettes. The states 
    targeted by the CORE algorithm are indicated by arrows together
    with their $SU(2)$ degeneracy. 
    \label{fig:2Plaquettes}
  }
\end{figure}

A next control step consists in calculating the spectrum of two 
coupled plaquettes, and one monitors which states are targeted by
the CORE algorithm. We show this spectrum in Fig.~\ref{fig:2Plaquettes}
along with the targeted states. We realize that the 16 states of our
tensor product basis cover almost all the low energy levels of the coupled
system. There are only two triplets just below the $S=2$ multiplet which are
missed.

\begin{figure}
  \includegraphics*[width=0.9\linewidth]{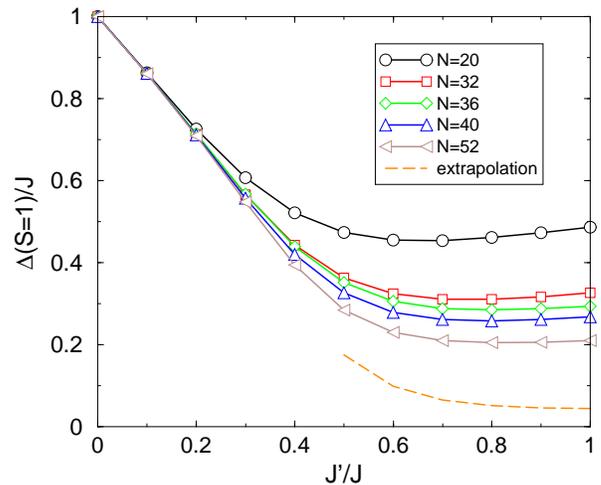}
  \caption{\label{fig:PlaquetteGap}
    Triplet Gap for effective system sizes between 20 and 52 sites,
    as a function of the interplaquette coupling $J'/J$.
    For $J'/J\ge 0.5$ a simple extrapolation in $1/N$ is also displayed.
    These results compare very well with ED results on
    the original model [\protect{\onlinecite{VoigtED}}].
  }
\end{figure}

In a first application we calculate the spin gap for different system sizes 
and couplings $J'/J$. The results shown in Fig.~\ref{fig:PlaquetteGap}
indicate a reduction of the spin gap for increasing $J'/J$. We used
a simple finite size extrapolation in $1/N$ in order to assess the 
closing of the gap. The extrapolation levels off to a small
value for $J'/J\ge0.6$. The appearance of a small gap in this 
known gapless region is a feature already present in ED calculation
of the original model \cite{VoigtED}, and therefore {\em not} an artefact 
of our method. It is rather obvious that the triplet gap 
is not a very accurate tool to detect the quantum phase transition 
within our numerical approach. We will see later that order parameter
susceptibilities are much more accurate.

\begin{figure}
  \includegraphics*[width=0.8\linewidth]{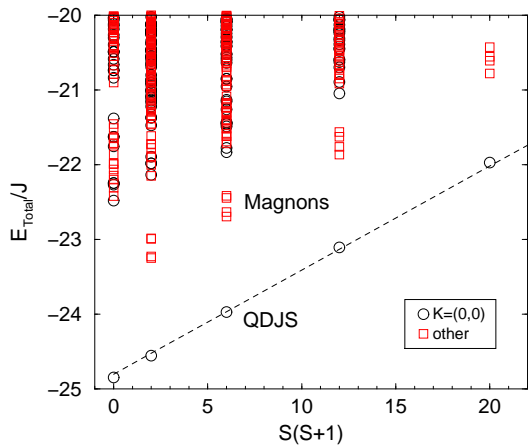}
  \caption{\label{fig:TowerSquare}
    Tower of states obtained with a range-2 CORE Hamiltonian on an
    effective $N=36$ square lattice (9-site CORE cluster)
    in different reduced momentum sectors. The tower of states 
    is clearly separated from the decimated magnons and the rest
    of the spectrum.
  }
\end{figure}

It is well known that the square lattice $(J'/J=1)$ is N\'eel ordered. One 
possibility to detect this order in ED is to calculate the so-called
{\em tower of excitation}, i.e. the complete spectrum as a function
of $S(S+1)$, $S$ being the total spin of an energy level. In the case of
standard collinear N\'eel order a prominent feature is an alignment of the 
lowest level for each $S$ on a straight line, forming a so called
``Quasi-Degenerate Joint States'' (QDJS) ensemble \cite{TowerTriangular},
which is clearly separated from the rest of the spectrum on a finite size 
sample. We have calculated the tower of states within the CORE approach
(Fig.~\ref{fig:TowerSquare}). Due to the truncated Hilbert
space we cannot expect to recover the entire spectrum.
Surprisingly however the CORE tower of states successfully reproduces the general 
features observed in ED calculations of the same model \cite{TowerEDSquare}:
(a) a set of QDJS with the correct degeneracy and quantum numbers 
(in the folded Brillouin zone);
(b) a reduced number of magnon states at intermediate energies, 
both set of states rather well separated from the high energy part of the 
spectrum. While the QDJS seem not to be affected by the CORE decimation 
procedure, clearly some the magnon modes get eliminated by the basis 
truncation. 

In order to locate the quantum phase transition from the paramagnetic, gapped 
regime to the N\'eel ordered phase, a simple way to determine the onset of long
range order is desireable. We chose to directly couple the order parameter to 
the Hamiltonian and to calculate generalized susceptibilities by deriving the 
energy with respect to the external coupling. This procedure is detailed in appendix 
\ref{sec:ObservablesNumericalCoreMethod}. Its simplicity relies on the fact that only
eigenvalue runs are necessary. Similar approaches have been used so far in ED and QMC 
calculations \cite{calandra00,capriotti}.

\begin{figure}
  \centerline{\includegraphics*[width=0.9\linewidth]{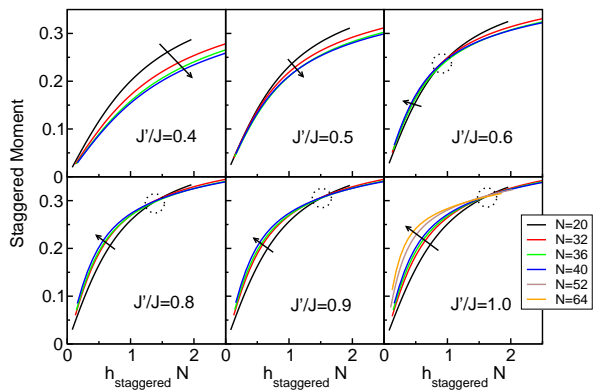}}
  \caption{
    Staggered moment per site as a function of the rescaled applied staggered
    field for the plaquette lattice and different values of $J'/J$.
    Circles denote the approximate crossing point of curves for different system
    sizes. We take the existence of this crossing as a phenomenological 
    indication for the presence of N\'eel LRO. In this way the phase transition
    is detected between $0.5<J'_c/J<0.6$, consistent with previous estimates.
    The arrows indicate curves for increasing system sizes.
    \label{fig:MstaggeredPlaquette}
  }
\end{figure}

Our results in Fig.~\ref{fig:MstaggeredPlaquette} show the evolution
of the staggered moment per site in a rescaled external staggered field for different
inter-plaquette couplings $J'$ and different system sizes (up to $8\times8$
lattices). We note the appearance of an approximate crossing of the curves for
different system sizes, once N\'eel LRO sets in. This approximate crossing relies
on the fact that the slope of $m_L(hN)$ diverges with increasing $N$ in the ordered
phase in our case \cite{capriotti}. We  then consider this crossing feature as an 
indication of the phase transition and obtain a value of the critical point
$J_c/J = 0.55 \pm 0.05$. This estimate is in good agreement with previous studies 
using various methods \cite{KogaSeriesPlaquette,AMLShastry,VoigtED}.
We have checked the present approach by performing the same steps on the
two leg ladder discussed in section \ref{sec:LadderGeometries} and there was no long
range magnetic order present, as expected.

\subsection{\kag systems with half-integer spins}

In the past 10 years many efforts have been devoted to
understand the low energy physics of
the \kag antiferromagnet (KAF) for spins $1/2$ 
\cite{LeungElser,Lecheminant,Waldtmann,MilaPRL,MambriniMila}.
At the theoretical level,
the main motivation comes from the fact that this model is the only known
example of a two-dimensional Heisenberg spin liquid. Even though many questions
remain open, some very exciting low-energy properties of this system have 
emerged. Let us summarize them briefly:
(i) the GS is a singlet ($S=0$) and has no magnetic order. Moreover no
kind of more exotic ordering (dimer-dimer, chiral order, etc.) have been
detected using unbiased methods;
(ii) the first magnetic excitation is a triplet ($S=1$) separated from
the GS by a rather small gap of order $J/20$;
(iii) more surprisingly the spectrum appears as a continuum of states in  all spin sectors.
In particular the spin gap is filled with
an exponential number of singlet excitations: ${\cal N}_{\rm singlets} \sim
1.15^N$;
(iv) the singlet sector of the KAF can be very well reproduced by a
short-range resonating valence bond approach involving only nearest-neighbor dimers.

From this point of view, the spin $1/2$ KAF with its
highly unconventional low-energy physics appears to be a
very sharp test of the CORE method.
The case of higher half-integer spins $S=3/2, 5/2 \dots$ KAF is also
of particular interest, since it is covered by approximative experimental
realizations \cite{SCGO}. Even if some properties of
these experimental systems are reminiscent of the spin $1/2$ KAF
theoretical support is still lacking for higher spins due to the
increased complexity of these models.

In this section we discuss in detail the range-two CORE Hamiltonians for spin 1/2 
and 3/2 KAF considered as a set of elementary up-triangles with couplings $J$,
coupled by down-triangles with couplings $J'$ [see Fig.~\ref{fig:Lattices2D}~(b)].
The coupling ratio will be denoted by $\alpha=J'/J$. Before going any
further into the derivation of the CORE effective Hamiltonian let us start
with the conventional degenerate perturbation theory results. Note that in the
perturbative regime these two approaches yield the same effective
Hamiltonian.

As described in Appendix \ref{sec:TechnicalKagome}, the most general
two-triangle effective Hamiltonian involving only the two spin 1/2 degrees
of freedom on each triangle can be written in the following form:
\begin{eqnarray}
  \label{eq:general}
  {\cal H} =
  N a_0(\alpha)+
  \sum_{\langle i,j \rangle}&& (
  b_0(\alpha) \vec{\tau}_i.\vec{e}_{ij} \vec{\tau}_j.\vec{e}_{ij}
  \label{eqn:GeneralTriangleCOREHamiltonian}\\
  &&+ a_1(\alpha) \vec{\sigma}_i . \vec{\sigma}_j\nonumber \\
  &&+ b_1(\alpha) \vec{\sigma}_i . \vec{\sigma}_j (\vec{\tau}_i.\vec{e}_{ij})
  (\vec{\tau}_j.\vec{e}_{ij})\nonumber\\ 
  &&+ c_1(\alpha) \vec{\sigma}_i . \vec{\sigma}_j
  (\vec{\tau}_i.\vec{e}_{ij}+\vec{\tau}_j.\vec{e}_{ij}) )\nonumber.
\end{eqnarray}

In the spirit of Mila's approach~\cite{MilaPRL} for spin $1/2$ the first
order perturbative Hamiltonian in $\alpha$ can easily be extended to
arbitrary half-integer spin $S$:
\begin{eqnarray}
  {\cal H}^{\rm pert.} &=& \frac{\alpha}{9} \vec{\sigma}_i . \vec{\sigma}_j
  \label{eqn:PerturbativeTriangleCOREHamiltonian}\\
  &&
  \times \left ( 1-2(2S+1)\vec{\tau}_i.\vec{e}_{a} \right ) \left (
    1-2(2S+1)\vec{\tau}_j . \vec{e}_{b} \right )\nonumber
\end{eqnarray}
and the coefficients of (\ref{eq:general}) in the perturbative limit are
given as $a_1(\alpha)= \frac{\alpha}{9}$, 
$b_1(\alpha)= \frac{4\alpha}{9} (2S+1)^2$, 
$c_1(\alpha)=-\frac{2\alpha}{9} (2S+1)$ 
and $a_0(\alpha)=b_0(\alpha)=0$. 
%           (1.) S=1/2 and (2. S=3/2)
\subsubsection{Choice of the CORE basis}

\begin{figure}
  \centerline{\includegraphics*[width=0.9\linewidth]{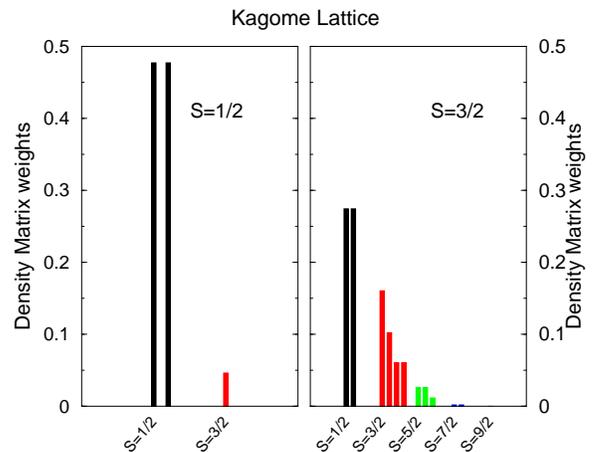}}
  \caption{
    Density matrix weights of the different total spin states in a triangle of a 
    12 site \kag cluster with $S=1/2$ and $S=3/2$ spins. These results are obtained
    for the homogeneous case $\alpha=1$.
    \label{fig:TriangleDensityMatrix}
  }
\end{figure}

As discussed in the previous paragraph we keep the two degenerate 
$S=1/2$ doublets on a triangle for the CORE basis. In analogy to the
the plaquette lattice we calculate the density matrix of 
a single triangle embedded in a 12 site \kag lattice for both spin $S=1/2$ and $S=3/2$,
in order to get information on the quality of the truncated basis.
The results displayed in Fig.~\ref{fig:TriangleDensityMatrix} show two different
behaviors: while the targeted states exhaust 95\% for the $S=1/2$ case, they 
cover only $\approx$ 55\% in the $S=3/2$ case. This can be considered a first 
indication that the range-two approximation in this basis might break down for  $S>1/2$ half
integer spin, while the approximation seems to work particularly well for $S=1/2$, thereby
providing independent support for the adequacy of the basis chosen in a related mean-field
study~\cite{MilaPRL}.

\begin{figure}
  \centerline{\includegraphics*[width=0.9\linewidth]{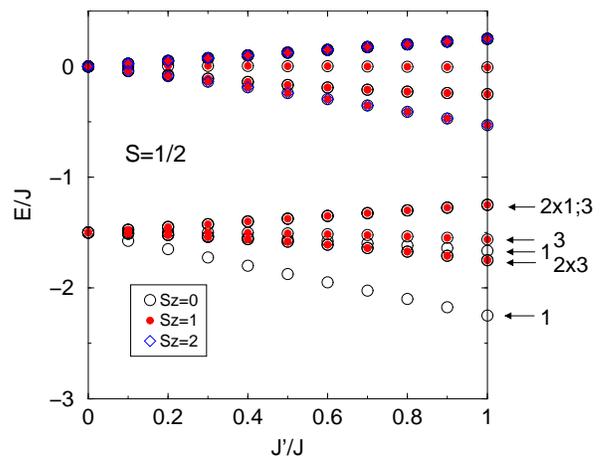}}
  \caption{\label{fig:SpectrumTwoTrianglesOneHalf}
    Spectrum of two coupled triangles in the \kag geometry
    with $S=1/2$ spins. The entire lowest band containing 16 states is
    successfully targeted by the CORE algorithm.
  }
\end{figure}

We continue the analysis of the CORE basis by monitoring the evolution of the 
spectra of two coupled triangles in the \kag geometry 
(c.f. Fig.~\ref{fig:TwoTriangles}) as a function of the
inter-triangle coupling $J'$, as well as the states selected by the range-two 
CORE algorithm. The spectrum for the spin $S=1/2$ case is shown in 
Fig.~\ref{fig:SpectrumTwoTrianglesOneHalf}. We note the presence of a clear gap 
between the 16 lowest states -- correctly targeted by the CORE algorithm -- 
and the higher lying bands. This can be considered an ideal case for the
CORE method. Based on this and the results of the density matrix we expect the
CORE range-two approximation to work quite well.

\begin{figure}
  \centerline{\includegraphics*[width=0.9\linewidth]{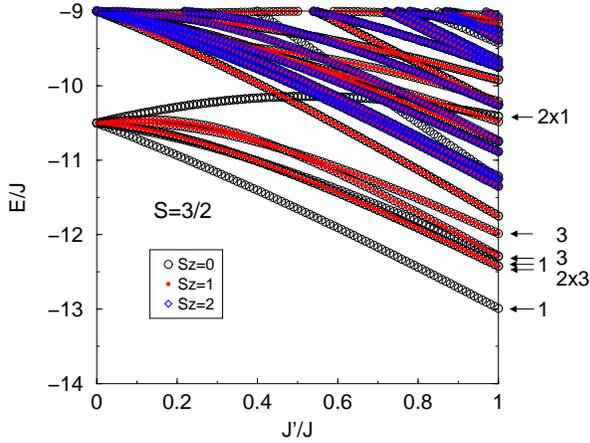}}
  \caption{\label{fig:SpectrumTwoTrianglesThreeHalf}
    Spectrum of two coupled triangles in the \kag geometry
    with $S=3/2$ spins. The 16 states targeted by the CORE algorithm
    are indicated by the arrows and their degeneracies.
  }
\end{figure}

We compare these encouraging results with the spectrum for the spin $S=3/2$ case
displayed in Fig.~\ref{fig:SpectrumTwoTrianglesThreeHalf}. Here the situation
is less convincing: very rapidly $(J'/J \gtrsim 0.45)$ the low energy states mix
with originally higher lying states and the CORE method continues to target two 
singlets which lie high up in energy when reaching $J'/J=1$. We expect this to be
a situation where the CORE method will probably not work correctly when restricted
to range-two terms only.

\begin{figure}
  \centerline{\includegraphics[width=0.9\linewidth]{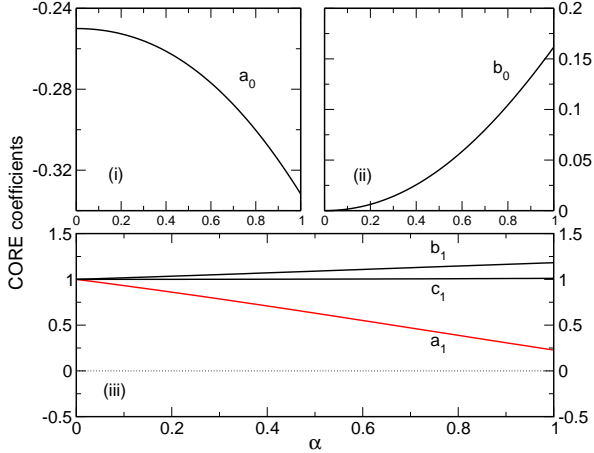}}
  \caption{\label{fig:CoreKagome12}
    Coefficients of the CORE range-two Hamiltonian for two coupled $S=1/2$ 
    triangles. The coefficients in panel (iii) have been divided by their
    values in the perturbative limit.}
\end{figure}

\begin{figure}
  \centerline{\includegraphics[width=0.9\linewidth]{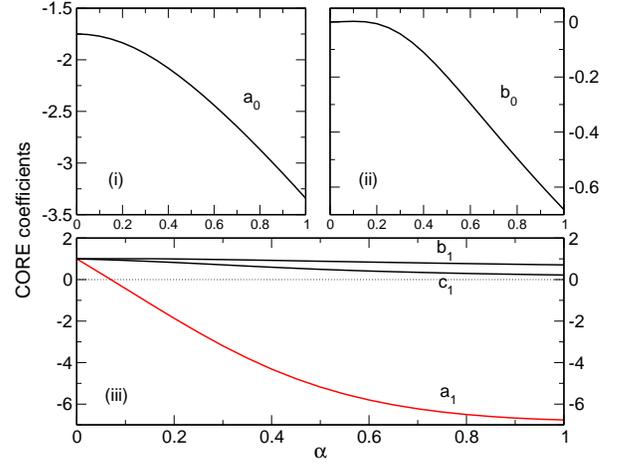}}
  \caption{\label{fig:CoreKagome32} 
    Coefficients of the CORE range-two Hamiltonian for two coupled $S=3/2$ 
    triangles. The coefficients in panel (iii) have been divided by their
    values in the perturbative limit.}
\end{figure}

Based on the two-triangle spectra shown above we used the CORE algorithm to
determine the coefficients of the general two-body
Hamiltonian [Eqn.~(\ref{eqn:GeneralTriangleCOREHamiltonian})]. For an independent derivation, 
see Ref.~[\onlinecite{Auerbach_Kagome}]. The coefficients
obtained this way are shown in Figs.~\ref{fig:CoreKagome12} and \ref{fig:CoreKagome32}
for $S=1/2$ and $S=3/2$ respectively. 
In the limit $\alpha \ll 1$
the coefficients can be obtained from the perturbative Hamiltonian 
[Eqn.~(\ref{eqn:PerturbativeTriangleCOREHamiltonian})]. There are two classes
of coefficients in both cases: $a_0$ and $b_0$ are zero in the perturbative
limit, i.e. they are at least second order in $\alpha$. The second class of
coefficients ($a_1$, $b_1$, $c_1$) are linear in $\alpha$. For improved visualisation
we have divided all the coefficients in the second class by their perturbative values.
In this way we observe in Fig.~\ref{fig:CoreKagome12} that coefficients $b_1$ and $c_1$
change barely with respect to their values in the perturbative limit. However $a_1$
has a significant subleading contribution, which leads to a rather large reduction
upon reaching the $\alpha=1$ point. It does however not change sign.

The situation for the $S=3/2$ case in Fig.~\ref{fig:CoreKagome32} is different: 
while the coefficients $b_1$ and $c_1$ decrease somewhat, it is mainly $a_1$ 
which changes drastically as we increase $\alpha$. Starting from 1 it rapidly goes
through zero ($\alpha\approx0.07$) and levels off to roughly -7 times the value
predicted by perturbation theory as one approaches $\alpha=1$. In this case it is
rather obvious that this coefficient will dominate the effective Hamiltonian. We
will discuss the implications of this behavior in the application to the $S=3/2$
\kag magnet below.

Let us note that the behavior of the $a_1$ coefficient is mainly due to a rather large
second order correction in perturbation theory. Indeed we find good agreement with the 
values obtained in the perturbative approach of Ref.~[\onlinecite{Raghu}].

\subsubsection{Simulations for $S=1/2$}

After having studied the CORE basis and the effective Hamiltonian at range two
in some detail, we now proceed to the actual simulations of the resulting model.
We perform the simulations for the standard \kag lattice, therefore $\alpha=1$.
We will calculate several distinct physical properties, such as the tower of 
excitations, the evolution of the triplet gap as a function of system size and
the scaling of the number of singlets in the gap. These quantities have been 
discussed in great detail in previous studies of the \kag $S=1/2$ 
antiferromagnet~\cite{LeungElser,Lecheminant,Waldtmann,MilaPRL,MambriniMila}. 

\begin{figure}
  \centerline{\includegraphics*[width=0.8\linewidth]{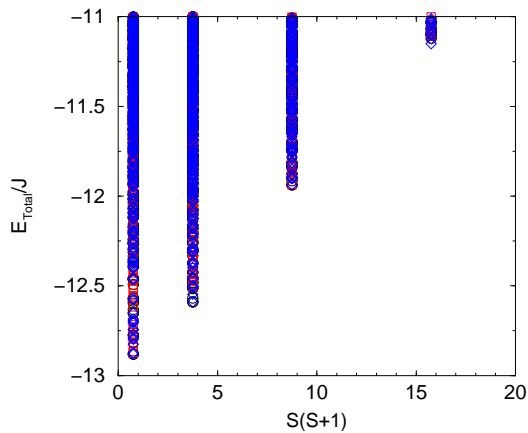}}
  \caption{\label{fig:TowerKagome}
    Tower of states obtained with a range-two CORE Hamiltonian on an
    effective $N=27$ \kag lattice (9-site CORE cluster).
    There is a large number of low-lying states in each $S$ 
    sector. The symbols correspond to different momenta.
  }
\end{figure}

First we calculate the tower of excitations for a \kag $S=1/2$ system on a 27 sites
sample. The data is plotted in Fig.~\ref{fig:TowerKagome}. The structure of
the spectrum follows the exact data of Ref.~[\onlinecite{Lecheminant}] rather closely; 
i.e there is no QDJS ensemble visible, a large number of $S=1/2$ states covering
all momenta are found below the first $S=3/2$ excitations and the spectrum is roughly 
bounded from below by a straight line in $S(S+1)$. Note that the tower of states we
obtain here is strikingly different from the one obtained in the N\'eel ordered 
square lattice case, see Fig.~\ref{fig:TowerSquare}.

\begin{figure}
  \centerline{\includegraphics*[width=0.80\linewidth]{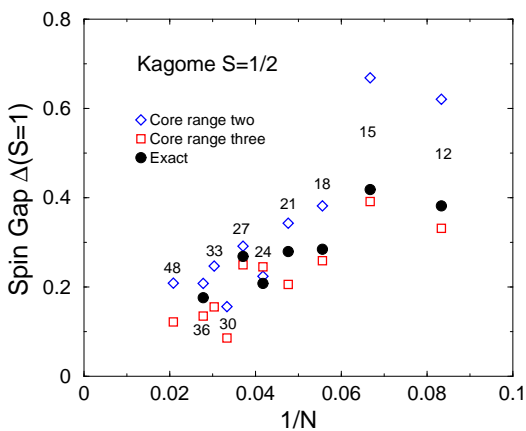}}
  \caption{\label{fig:CoreOneHalfGap}
    Spin gap of the \kag $S=1/2$ model on various samples, obtained with the
    CORE method (range-two and three). Exact diagonalization
    result are also shown for comparison where available.}
\end{figure}

Next we calculate the spin gap using the range-two CORE Hamiltonian. Results for
system sizes up to 48 sites are shown in Fig.~\ref{fig:CoreOneHalfGap},
together with ED data where available. In comparison we note two observations: (a)
the CORE range-two approximation seems to systematically overestimate the gap, 
but captures correctly the sample to sample variations. (b) the gaps of the
smallest samples (effective $N$=12,15) deviate strongly from 
the exact data. We observed this to be a general feature of very small clusters 
in the CORE approach. In order to improve the agreement with the ED data we calculated
the two CORE range-three terms containing a closed loop of triangles. The results obtained
with this extended Hamiltonian are shown as well in Fig.~\ref{fig:CoreOneHalfGap}.
These additional terms improve the gap data somewhat. We now find the CORE gaps to be mostly
smaller than the exact ones. The precision of the CORE gap data is
not accurate enough to make a reasonable prediction on the spin gap in the thermodynamic
limit. However we think that the CORE data is compatible with a finite spin gap.

\begin{figure}
  \centerline{\includegraphics*[width=0.75\linewidth]{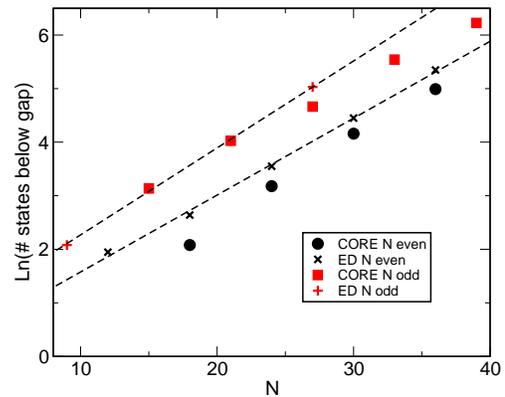}}
  \caption{\label{fig:KagomeOneHalfSinglets}
    Logarithm of the number of states within the magnetic gap. Results obtained 
    with the CORE range-two Hamiltonian. For comparison exact data obtained in 
    Refs.~[\protect{\onlinecite{Lecheminant}},\protect{\onlinecite{Waldtmann}}]
    are shown. The dashed lines are linear fits to the exact diagonalization data.
  }
\end{figure}

Finally we determine the number of nonmagnetic excitations within the magnetic gap
for a variety of system sizes up to 39 sites. Similar studies of this quantity in ED gave 
evidence for an exponentially increasing number of singlets in the gap
\cite{Lecheminant,Waldtmann}. We display our data in comparison to the exact results
in Fig.~\ref{fig:KagomeOneHalfSinglets}. While the precise numbers are not expected
to be recovered, the general trend is well described with the CORE results. For both even
and odd $N$ samples we see an exponential increase of the number of these nonmagnetic 
states. In the case of $N=39$ for example, we find 506 states below the first magnetic
excitation. These results emphasize again the validity of the two doublet basis for
the CORE approach on the \kag spin 1/2 system.

\subsubsection{Simulations for $S=3/2$}

We have also simulated the CORE Hamiltonian obtained above for $S=3/2$. 
While the energy per site is reproduced roughly, unfortunately the spectrum 
does not resemble an antiferromagnetic spin model, i.e. the groundstate is
polarized in the spin variables. This fact is at odds with preliminary exact
diagonalization data on the original $S=3/2$ model \cite{KagThreeHalf}. We 
therefore did not pursue the CORE study with this choice of the
basis states any further. Indeed, as suggested by the analysis of the density
matrix and by the evolution of the spectrum of two coupled triangles, we consider
this a breakdown example of a naive range-two CORE approximation. It is important
to stress that the method indicates its failure in various quantities throughout
the algorithm, therefore offering the possibility of detecting a possible breakdown.

As a remedy in the present case we have extended the basis states to include all
the $S=1/2$ and $S=3/2$ states on a triangle (i.e. keeping 20 out of 64 states).
Computations within this basis set are more demanding, but give a better agreement
with the exact diagonalization results. At the present stage we cannot decide whether
the breakdown of the 4 states CORE basis is related only the CORE method
or whether it implies that the \kag $S=1/2$ and $S=3/2$ systems do not belong to
the same phase.

\section{Conclusions}

We have discussed extensively the use of a novel numerical technique - the so-called numerical 
Contractor Renormalization (CORE) method - in the context of low-dimensional quantum magnetism. 
This method consists of two steps: (i) building an effective Hamiltonian
acting on the low-energy degrees of freedom of some elementary block; and (ii) studying this new model 
numerically on finite-size clusters, using a standard Exact Diagonalization or similar approach. 

Like in other real-space renormalization techniques the effective model usually contains longer range
interactions. The numerical CORE procedure will be most efficient provided the effective interactions 
decay sufficiently fast. We discussed the validity of this assumption in several cases. 

For ladder type geometries, we explicitely checked the accuracy of the effective models by
increasing the range of the effective interactions until reaching convergence. Both in the perturbative
regime and in the isotropic case, our results on a 2-leg ladder and a 3-leg torus are in good agreement 
with previously established results. This rapid convergence might be due to the small correlation length
that exists in these systems which both have a finite spin gap.

In two dimensions, we have used the density matrix as a tool to check whether the restricted basis gives
a good enough representation of the exact states.  When this is the case, as for the plaquette lattice or
the $S=1/2$ \kag lattice, the lowest order range-two effective Hamiltonian gives semi-quantitative results, 
even away from any perturbative regime. For example we can successfully describe the plaquette lattice,
starting from the decoupled plaquette limit through the quantum phase transition to the N\'eel ordered
state at homogeneous coupling. Furthermore we can also reproduce many  aspects of the exotic low-energy
physics of the $S=1/2$ \kag lattice.

Therefore within the CORE method, we can have both the advantage of working in a strongly reduced subspace 
and not being limited to the perturbative regime in certain cases.

We thus believe that the numerical CORE method can be used systematically to explore possible ways of
generating low-energy effective Hamiltonians. An important field is for example the doped frustrated
magnetic systems, where it is not easy to decide which states are important in a low-energy 
description, and therefore the density matrix might be a helpful tool.

\appendix

\section{Density Matrix}
\label{sec:DensityMatrix}

In this appendix we introduce the density matrix of a basic building block in a larger
cluster of the fully interacting problem as a diagnostic tool to validate or invalidate a 
particular choice of retained states on the basic building block in the CORE approach.

In previous applications of the CORE method, the choice of the states kept relied mostly
on the spectrum of an isolated building block. While this usually gives reasonable results
it is not a clear {\em a priori } where to place the cut-off in the spectrum.

The density matrix of a ``system block'' embedded in a larger ``super block'' forms a key
concept in the Density Matrix Renormalization Group (DMRG) algorithm invented by S.R.~White
in 1992~\cite{DMRG} and is at the heart of its success. Based on this and related ideas
\cite{DMPhonons} we propose to monitor the density matrix of the basic building block 
embedded in a larger cluster and to retain these states exhausting a large fraction of the
density matrix weight.

Consider now a subsystem $\mathcal{A}$ embedded in a larger system $\mathcal{B}$. Suppose
that the overall system $\mathcal{B}$ is in state $|\Psi\rangle$ (e.g. the groundstate). 
We write the wavefunction as:
\begin{equation}
|\Psi\rangle= \sum_{a,b}\ \psi_{a,b}\ |a\rangle\otimes|b\rangle,
\end{equation}
where the sum index $a$ runs over all states in $\mathcal{A}$ and index $b$ over all states in
$\mathcal{B} \setminus \mathcal{A}$.
The density matrix $\rho^\mathcal{A}$ of the subsystem $\mathcal{A}$ is then
defined as
\begin{equation}
\rho^\mathcal{A}_{a,a'}=\sum_{b} \psi_{a,b} \psi_{a',b}^*
\end{equation}
The eigenvalues of $\rho^\mathcal{A}$ denote the probability of finding a certain
state $a$ in $\mathcal{A}$, given the overall system in state $|\Psi\rangle$.

Practically we calculate the groundstate of the fully interacting system on a medium
size cluster by exact diagonalization, and then obtain the density matrix of a basic
building block, e.g. a four site plaquette. The density matrix of a building block is
a rather local object, so we expect that results on intermediate size clusters are 
already accurate on the percent level. The density matrix spectra shown in
Figs.~\ref{fig:PlaquetteDensityMatrix} and \ref{fig:TriangleDensityMatrix} have been 
obtained in this way. In the models considered, a density matrix weight of the
retained states of at least 90\% yielded reasonable results within a range-two CORE 
approximation. It is possible to allow for a lower overall weight, at the expense 
of increasing the range of the CORE interactions.

\section{Observables in the numerical CORE method}
\label{sec:ObservablesNumericalCoreMethod}

The calculation of observables beyond simple energy related quantities
is not straightforward within the CORE method, as the observables need
to be renormalized like the Hamiltonian in the first place
\cite{core-bosonfermion,dynamicCORE}.

A somewhat simpler approach for measurements of symmetry breaking order
parameters consists in adding a small symmetry breaking field to the 
Hamiltonian (for a review see Ref.~[\onlinecite{capriotti}]).

Let us denote ${\cal \hat O}$ the extensive symmetry breaking operator,
such that the order parameter is related to its GS average value 
$m=1/N \langle \psi_0|{\cal \hat O}|\psi_0\rangle$. The occurence of a 
symmetry broken phase can be detected by adding this 
operator to the Hamiltonian~:
\begin{equation}
  {\cal H}(\delta) ={\cal H} - \delta {\cal \hat O}
\end{equation}
Since on a finite-size lattice the order parameter vanishes by symmetry for
$\delta=0$, the ground-state energy per site varies quadratically for small 
$\delta$~: 
$$e(\delta)\simeq e_0 -\frac{1}{2}\chi_0 \delta^2,$$
where $\chi_0$ is termed the corresponding generalized susceptibility.
In that way the second derivative of the energy with respect to $\delta$ at
$\delta=0$ offers one possibility to detect a finite order parameter in the 
thermodynamic limit \cite{capriotti}. 

We found that another possibility to conveniently track the presence of a 
finite order parameter is to measure directly $m(\delta)$ in finite field
$$ m(\delta)=\langle\Psi_\delta|{\cal \hat O}|\Psi_\delta\rangle=
\mbox{d}e(\delta)/\mbox{d}\delta$$ by the Hellmann-Feynman theorem. When
plotting $m(\delta)$ as a function of the rescaled field $N \delta$ for 
various system sizes we observe an approximate crossing of the curves
if there is a finite order parameter and no crossing in the absence of
the order parameter.

\section{Gauge invariance on half-integer spins \kag like systems}
\label{sec:TechnicalKagome}

In this appendix, we discuss half-integer spin Hamiltonians with 
triangles as the unit cell. The ground state manifold of each unit cell
is generated by the four degenerate lowest states that can
be built out of $3$ half-integer $S$ spins, namely the four 
$S_{\rm tot}=1/2$ states. 
The idea of selecting these states as a starting point to describe
the whole system low energy properties was originally introduced by
Subrahmanyam for $S=1/2\ $~\cite{Subrahmanyam} on the \kag lattice and
later used by Mila~\cite{MilaPRL}. More recently it was reintroduced by
Raghu {\it et al}~\cite{Raghu} for arbitrary half-integer $S$ in the context
of a chain of triangles. All these approaches are pertubative and state that
the triangle couplings $J$ is much larger than the inter-triangle one $J'$.

Here we would like to discuss some general
properties of any effective Hamiltonian that can be derived either by
perturbative methods or more sophisticated ones such as CORE. In particular,
we would like to point out that a gauge invariance appears as a direct
consequence of the state selection.

\begin{figure}
  \centerline{\includegraphics[width=0.7\linewidth]{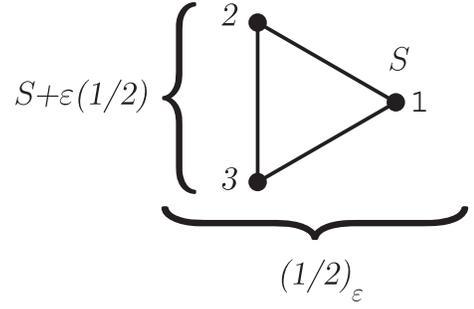}}
  \caption{\label{fig:OneTriangle} Definition of chirality $\varepsilon$ (see
    text for details).}
\end{figure}

To be more specific, let us label $1$,$2$,$3$ the sites of the triangle (see
Fig.~\ref{fig:OneTriangle}). In order to build a total spin $1/2$ out of the
three $S$, spins $2$ and $3$ couple into a
$S+\varepsilon(1/2)$ with $\varepsilon=\pm 1$. The coupling with the remaining
site $1$ produces a spin $1/2$ with chirality $\varepsilon=\pm 1$. Note that
this definition of chirality is equivalent to Eqs.~(\ref{eqn:TorusChirality})
for spin $S=1/2$ up to a global unitary transform which is just a
redefinition of the chirality quantification axis.

In the following, the four  selected spin-chirality states on a triangle $i$
will be denoted as $\vert \vert \varepsilon_i , \nu_i \rangle \rangle$. 
These states are the eigenstates of the $z$ components of spin
$\vec{\sigma}$ and chirality $\vec{\tau}$ (both are spin 1/2 like operators)
with $\tau_z \vert \vert
\varepsilon_i , \nu_i \rangle \rangle = (\varepsilon_i/2) \vert \vert
\varepsilon_i , \nu_i \rangle \rangle$ and $\sigma_z \vert \vert
\varepsilon_i , \nu_i \rangle \rangle = \nu_i \vert \vert
\varepsilon_i , \nu_i \rangle \rangle$.

\begin{figure}
\centerline{\includegraphics[width=0.7\linewidth]{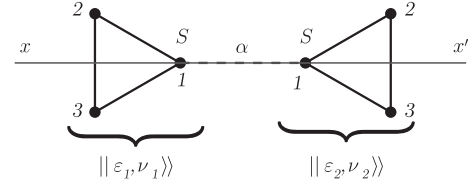}}
\caption{\label{fig:TwoTriangles} The two-triangle problem. $\alpha$ is the
  coupling ratio $J'/J$.}
\end{figure}

Let us now turn to the two-triangle problem. As it can be seen
in Fig.~\ref{fig:TwoTriangles}, the Hamiltonian is invariant under 
reflections with respect to the $(xx')$ axis. Moreover, the reflection
can be taken independently on each triangle. As a consequence, both chiralities 
($\tau_i^z$ and $\tau_j^z$) are conserved by 
the effective Hamiltonian and the $\tau$ part is of the form $1+a
(\tau_i^z+\tau_j^z)+ b \tau_i^z \tau_j^z$. For any fixed value of
$(\varepsilon_i,\varepsilon_j)$, the total spin of the system is conserved and thus
the spin part is $SU(2)$ invariant. As a conclusion the most general
two-triangle Hamiltonian allowed is of the form:
\begin{equation*}
(\vec{\sigma}_i \cdot \vec{\sigma}_j + c )(1+a
(\tau_i^z+\tau_j^z)+ b \tau_i^z \tau_j^z).
\end{equation*}

{\it Gauge transformation:} The form of the above Hamiltonian
is the consequence of the particular choice we made for labeling the sites
of the triangle (see Fig.~\ref{fig:TwoTriangles}): site $1$ of triangle $t_1$
couples to site $1$ of triangle $t_2$. Although this gauge was convenient for
the calculation, in general this choice can not be made simultaneously on
all couples of triangles of the lattice. So, it is essential to derive the form
of the Hamiltonian in a generic situation where site $i=1,2,3$ of triangle
$t_1$ couples to site $j=1,2,3$ of triangle $t_2$. 

\begin{figure}
  \centerline{\includegraphics[width=0.9\linewidth]{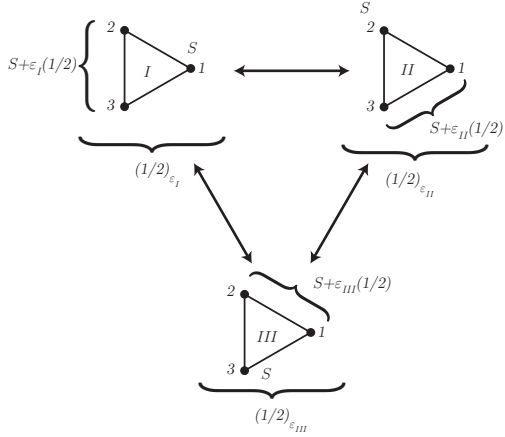}}
  \caption{\label{fig:3j_symbols} 
    Three ways of coupling the three spins
    $S$ on a triangle into a total spin $1/2$ state. Each construction
    is related to the two others by the 3$j$ symbols (see text).
  }
\end{figure}

The unitary transformations involved in the redefinition of the coupling
sequence (see Fig.~\ref{fig:3j_symbols}) are covered by the $3j$ symbols of
elementary quantum mechanics. The problem of 3 half-integer spins $S$ coupled 
into a total spin $1/2$ occurs to be particularly simple and independent of
$S$. The form of the general effective Hamiltonian then reads:
\begin{eqnarray*}
{\cal H}_{ij}^{a,b}(\alpha) &=& \left ( \vec{\sigma}_i \cdot \vec{\sigma}_j +
c(\alpha) \right )\\
&& \times
 [ 1+a(\alpha)(\vec{\tau}_i\cdot\vec{e}_{a}+\vec{\tau}_j\cdot\vec{e}_{b})\\
& &\quad + b(\alpha)
(\vec{\tau}_i\cdot\vec{e}_{a})(\vec{\tau}_j\cdot\vec{e}_{b})],
\end{eqnarray*}
where $\vec{e}_a$, $a=1,2,3$ are three coplanar normalized vectors in a $120^\circ$
configuration (for example, $\vec{e}_1=(0,1)$,  $\vec{e}_2=(-\sqrt{3}/2,-1/2)$ and
$\vec{e}_3=(\sqrt{3}/2,-1/2)$ in the $x-z$ plane) and $a$, $b$ are the labels
of the original spins coupling triangles $t_i$ and $t_j$.

{\it The \kag lattice:} In the particular geometry of the \kag
lattice [see Fig.~\ref{fig:Lattices2D} (b)], each triangular unit cell is 
coupled to six other triangular cells, each corner being coupled twice. 
As a consequence, for each cell the
contribution involving only $\vec{\tau}_i\cdot\vec{e}_{\alpha}$ factorizes
into $2 \vec{\tau}_i \cdot (\vec{e}_1+\vec{e}_2+\vec{e}_3) = 0$. The
corresponding terms are then not relevant in the Hamiltonian and thus 
we denote the most general two-triangle Hamiltonian for the \kag lattice as: 
\begin{eqnarray*}
  {\cal H}^ =
  N a_0(\alpha)+
  \sum_{\langle i,j \rangle}&& [
  b_0(\alpha) \vec{\tau}_i\cdot\vec{e}_{ij} \vec{\tau}_j\cdot\vec{e}_{ij}\\
  &&+ a_1(\alpha) \vec{\sigma}_i \cdot \vec{\sigma}_j \\
  &&+ b_1(\alpha) \vec{\sigma}_i \cdot \vec{\sigma}_j (\vec{\tau}_i\cdot\vec{e}_{ij})
  (\vec{\tau}_j\cdot\vec{e}_{ij})\\ 
  &&+ c_1(\alpha) \vec{\sigma}_i \cdot \vec{\sigma}_j
  (\vec{\tau}_i\cdot\vec{e}_{ij})+(\vec{\tau}_j\cdot\vec{e}_{ij}) ]
\end{eqnarray*}
which is the form used in the text. 

% Acknowledgments
\acknowledgments
We thank F.~Alet, A.~Auerbach, F. Mila and D.~Poilblanc for fruitful discussions.
Furthermore we are grateful to F.~Alet for providing us QMC data.
We thank M.~K\"orner for his very useful \verb:Mathematica: spin notebook.
A.L.~acknowledges support from the Swiss National Fund.
We thank IDRIS (Orsay) and the CSCS Manno for allocation of CPU time.

% Bibliography

\end{document}